# Single-Particle Spectrum of Pure Neutron Matter

Khalaf Gad[1*] and Hesham Mansour[2]

[1]*Physics Department, Faculty of Science, Sohag University, Sohag, Egypt*
[2]*Physics Department, Faculty of Science, Cairo University, Giza, Egypt*

We have calculated the self-consistent auxiliary potential effects on the binding energy of neutron matter using the Brueckner–Hartree–Fock approach by adopting the Argonne V18 and CD-Bonn potentials. The binding energy with the four different choices for the self-consistent auxiliary potential is discussed. Also, the binding energy of neutron matter has been computed within the framework of the self-consistent Green's function approach. We also compare the binding energies obtained in this study with those obtained by various microscopic approaches. It is found that the use of the continuous choice tends to give binding energies about 2–4 MeV larger than the gap choice at $k_F = 1.8$ fm$^{-1}$. In the case of symmetric nuclear matter this difference is larger.

## 1. Introduction

The equation of state (EOS) of neutron matter[1–6] plays a central role in astrophysics, for problems ranging from the structure of neutron stars[7,8] and neutron star mergers[9,10] to core-collapse supernovae.[11] The mass of a neutron star depends mainly on the EOS of neutron matter up to a density $\rho \approx 4\rho_0$. It has been argued that different choices for the self-consistent auxiliary potential have an effect on the convergence rate of hole-line expansion and can produce energy shifts in the calculated total binding energy.[12,13] Also under both the gap and continuous choices, the three-body force (TBF) plays an important role in determining the high-density behavior of symmetry energy, and its effect leads to a strong stiffening of the symmetry energy at high densities.[14–17]

Baldo and Fiasconaro[18] studied the single particle spectrum and binding energy of symmetric nuclear matter. They found that the parabolic approximation for the single-particle potential $U(k)$ in the self-consistent Brueckner–Hartree–Fock (BHF) scheme introduces an uncertainty of 1–2 MeV near the saturation density; therefore, it cannot be used in accurate calculations. In Brueckner–Hartree–Fock calculations of nuclear matter, the self-consistent single-particle potential is strongly momentum-dependent.

Frick et al.[19] studied the sensitivity of BHF approximation for the many-body system of symmetric nuclear matter with respect to the exact treatment of the propagator in the Bethe–Goldstone equation. One finds that the precise treatment of the Pauli operator together with a single-particle spectrum based on the real part of the self-energy for the hole and particle states yields a binding energy per nucleon that is larger by a nonnegligible amount than those obtained in standard approximation schemes.

Wang and Zuo[20] studied the TBF effects on the properties of nuclear matter under the gap and continuous choices within the BHF approach. They found that the TBF provides a strong repulsive effect on the equation of state of nuclear matter at high densities for both choices. The saturation point turns out to be much closer to the empirical value when the continuous choice is adopted. Also it is seen that, there is a weak dependence on the choice of the auxiliary single-particle potential, as determined using the Argonne V18 and Bonn B two-body potentials.

Gandolfi et al.[17] studied the equation of state of pure neutron matter using quantum Monte Carlo (QMC) methods.

In recent years, a QMC technique, called auxiliary field diffusion Monte Carlo (AFDMC), has been developed to study large pure neutron systems with the same accuracy as Green's function Monte Carlo (GFMC). The GFMC technique has been hugely successful in calculating the ground-state properties of nuclei up to 12 nucleons using realistic nuclear Hamiltonians with local two- and TBF's.

It is well known that, in symmetric nuclear matter the TBF is required to obtain the correct saturation point. The inclusion of the TBF was considered by many authors, particularly in Ref. 21 where the BHF results for both neutron matter and symmetric nuclear matter were implemented by the inclusion of the Urbana TBF.[22,23]

In this work, we will extend the analysis to pure neutron matter EOS, which is crucial for neutron star studies. The effect of using different single-particle potentials is investigated. In order to analyze the dependence of the results on the nuclear interaction, two different realistic and accurate two-body forces are considered, Argonne V18[24] and CD-Bonn,[25] which give quite different EOS's. Moreover, the results are compared with those of various many-body approaches.

This paper is organized as follows. After the introduction, we briefly review the general picture in Sect. 2. The results and discussions are presented in Sect. 3. Conclusions are given in Sect. 4.

## 2. Theoretical Background

In this section, we briefly illustrate how to evaluate, in the BHF[12,13,26–28] and self-consistent Green's function (SCGF)[29–34] approximations, the energy per particle of a system of nucleons. Our calculation of the energy per particle starts with the construction of a $G$-matrix, which effectively describes the interaction between two nucleons. The $G$-matrix is obtained by solving the Bethe–Brueckner–Goldstone (BBG) equation

$$G(\omega) = v_{NN} + v_{NN} \sum_{k_1 k_2} \frac{|k_1 k_2\rangle Q_{k_1,k_2} \langle k_1 k_2|}{\omega - \varepsilon_{k_1} - \varepsilon_{k_2}} G(\omega), \quad (1)$$

where $\omega$ is the starting energy of the interacting nucleon, $v_{NN}$ is the free nucleon–nucleon (NN) interaction, and $Q$ is the Pauli operator that prevents scattering into occupied states. On the basis of the $G$-matrix value, one can calculate the single-particle energy of a nucleon with the momentum $k$ using

$$\varepsilon_k = \frac{\hbar^2 k^2}{2m_\text{n}} + U(k,\rho), \quad (2)$$

where the single-particle potential $U(k)$ represents the average field experienced by the nucleon due to its interaction with other nucleons in the system. According to Jeukenne et al.[35] the real part of self-energy represents the single-particle potential for the particle and hole states. Under the continuous choice, the auxiliary potential is given by

$$U(k,\rho) = \text{Re}\,\Sigma_{k' \leq k_\text{F}} \langle kk'|G(\rho,\omega = \varepsilon_k + \varepsilon_{k'})|kk'\rangle_A, \quad (3)$$

where the subscript $A$ indicates the antisymmetrization of the matrix element. In this scheme, the only input quantity we need is the bare NN interaction $v$ in the Bethe–Goldstone equation (1).

In order to obtain such a self-consistent solution of the BHF equations, one often assumes a quadratic dependence of the single-particle energy on the momentum of the neutron in the form

$$\varepsilon_k \approx \frac{\hbar^2 k^2}{2m_\text{n}^*} + C, \quad (4)$$

where $m_\text{n}^*$ is the effective mass for neutron matter and $C$ is a constant.

In pure neutron matter (PNM), only partial waves with a pair of interacting nucleons coupled to isospin $T = 1$ contribute to the calculation of the $G$-matrix value. Owing to the antisymmetry of matrix elements, only partial waves with even values for the sum $L + S$, like $^1S_0$, $^3P_0$, etc. are considered.[36] In the case of symmetric nuclear matter (SNM), other partial waves such as $^3S_1$–$^3D_1$ and $^1P_1$ contribute.

Once a self-consistent solution of Eqs. (1) and (3) is obtained, the energy per particle can be easily calculated using

$$\frac{E}{A} = \frac{3}{5}\frac{\hbar^2 k_\text{F}^2}{2m_\text{n}} + \frac{1}{2\rho}\text{Re}\,\Sigma_{k,k' \leq k_\text{F}}\langle kk'|G(\rho,\omega = \varepsilon_k + \varepsilon_{k'})|kk'\rangle_A. \quad (5)$$

One of the drawbacks of the BHF approach is the fact that it does not provide consistent results from the point of view of thermodynamics, i.e., it is not in agreement with the fulfillment of the Hugenholtz van Hove theorem. This is due to the fact that the BHF approximation does not consider the propagation of particle and hole states on equal footing. An extension of the BHF approach that obeys this symmetry is the self-consistent Green's function (SCGF) method using the so-called $T$-matrix approximation. In recent years, techniques that allow us to evaluate the solution of the SCGF equations for microscopic NN interactions[37,38] have been developed. Those calculations demonstrate that, in the case of realistic NN interactions, the contribution of particle–particle ladders dominates the contribution of the corresponding hole–hole propagation terms. This justifies the use of the BHF approach and a procedure that goes beyond BHF and accounts for hole–hole terms in a perturbative way. This leads to the modification of the self-energy in the BHF approximation by adding a hole–hole term of the form[19,39]

$$\Sigma^{2\text{h}1\text{p}}(k,\omega)$$
$$= \int_{k_\text{F}}^\infty d^3p \int_0^{k_\text{F}} d^3h_1\, d^3h_2 \frac{\langle k,p|G|h_1,h_2\rangle_A^2}{\omega + \varepsilon_p - \varepsilon_{h1} - \varepsilon_{h2} - i\eta}. \quad (6)$$

The quasi-particle energy for the extended self-energy can be defined as

$$\varepsilon_{ki}^\text{qp} = \frac{\hbar^2 k^2}{2m_\text{n}} + \text{Re}[\Sigma^\text{BHF}(k,\omega = \varepsilon_k^\text{qp}) + \Sigma^{2\text{h}1\text{p}}(k,\omega = \varepsilon_k^\text{qp})]. \quad (7)$$

Accordingly, the Fermi energy is obtained by evaluating the latter equation at the Fermi momentum $k = k_\text{F}$:

$$\varepsilon_\text{F} = \varepsilon_{k_\text{F}}^\text{qp}. \quad (8)$$

Assuming that the self-energy $[\Sigma(k,\omega) = \Sigma^\text{BHF}(k,\omega) + \Sigma^{2\text{h}1\text{p}}(k,\omega)]$ for a nucleon with the momentum $k$ and energy $\omega$ in infinite nuclear matter is given, the Dyson equation leads to a single-particle Green's function of the form

$$g(k,\omega) = \frac{1}{\omega - \dfrac{\hbar^2 k^2}{2m} - \Sigma(k,\omega)}. \quad (9)$$

If one compares this solution with the general Lehmann representation

$$g(k,\omega) = \lim_{\eta \to 0}\left(\int_{-\infty}^{\epsilon_\text{F}} d\omega' \frac{S_\text{h}(k,\omega')}{\omega - \omega' - i\eta} + \int_{\epsilon_\text{F}}^\infty d\omega' \frac{S_\text{p}(k,\omega')}{\omega - \omega' + i\eta}\right), \quad (10)$$

one can easily identify the spectral functions $S^\text{h}(k,\omega)$ and $S^\text{p}(k,\omega)$ for hole and particle strengths, respectively, to be given by

$$S^{\text{h}(\text{p})}(k,\omega) = \pm\frac{1}{\pi}\frac{\text{Im}\,\Sigma(k,\omega)}{[\omega - \hbar^2 k^2/2m - \text{Re}\,\Sigma(k,\omega)]^2 + [\text{Im}\,\Sigma(k,\omega)]^2}, \quad (11)$$

where the plus and minus signs on the left-hand side of this equation refers to the case of the hole ($h, \omega < \varepsilon_{\text{F}i}$) and particle ($p, \omega > \varepsilon_{\text{F}i}$) states, respectively.

In the SCGF approach, the particle states ($k > k_\text{F}$), which are missing in the BHF energy sum rule of Eq. (5), do contribute according to the energy sum rule[40]

$$\frac{E}{A} = \frac{\int d^3k \int_{-\infty}^{\varepsilon_\text{F}} d\omega\, S_\tau^\text{h}(k,\omega)\frac{1}{2}\left(\dfrac{\hbar^2 k^2}{2m} + \omega\right)}{\int d^3k\, n(k)}. \quad (12)$$

Equation (12) shows the link between the energy of the system and the hole spectral function, $S^\text{h}(k,\omega)$.

## 3. Results and Discussion

The binding energies of neutron matter calculated in the BHF approximation using Argonne V18 and CD-Bonn potentials are shown in Figs. 1 and 2, respectively, as a function of the Fermi momentum $k_\text{F}$ for single-particle potentials.

Four different choices have been adopted in the BHF calculations: The first one is the continuous spectrum with free energies below and above the Fermi surface plus a constant shift of particle potential above the Fermi surface (SPE 1).
The second one is the conventional choice for single-particle energy: the bound energy below the Fermi surface and free the energy above the Fermi surface (i.e., gap) (SPE 2).
The third one is the continuous spectrum with a bound energy below the Fermi surface and continuous continuation above the Fermi surface up to $U = 0$; free energies after (SPE 3).
The fourth one is continuous forever (SPE 4).

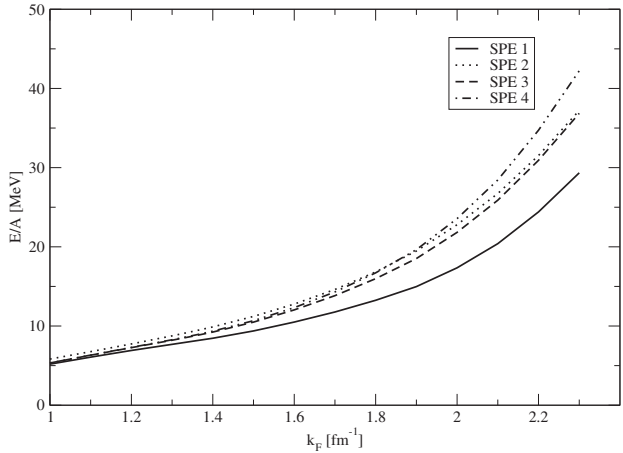

**Fig. 1.** Equation of state of neutron matter obtained using single-particle potentials, as described in the text using Argonne V18 potential.

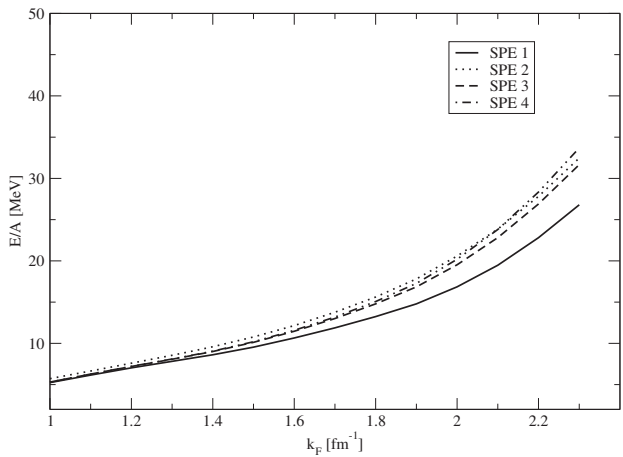

**Fig. 2.** Same as Fig. 1 but using the CD-Bonn potential.

**Table I.** Neutron matter binding energy per neutron (in MeV) as a function of Fermi momentum, $k_F$, for the Argonne V18 potential, as obtained by BHF calculation for different single-particle potentials.

| $k_F$ (fm$^{-1}$) | E/A SPE 1 | SPE 2 | SPE 3 | SPE 4 |
|---|---|---|---|---|
| 1 | 5.186 | 5.829 | 5.331 | 5.334 |
| 1.1 | 6.069 | 6.761 | 6.314 | 6.330 |
| 1.2 | 6.916 | 7.735 | 7.268 | 7.301 |
| 1.3 | 7.684 | 8.755 | 8.204 | 8.264 |
| 1.36 | 8.150 | 9.420 | 8.810 | 8.896 |
| 1.4 | 8.460 | 9.886 | 9.238 | 9.350 |
| 1.5 | 9.380 | 11.214 | 10.492 | 10.674 |
| 1.6 | 10.499 | 12.772 | 12.035 | 12.323 |
| 1.7 | 11.772 | 14.616 | 13.839 | 14.305 |
| 1.8 | 13.250 | 16.829 | 15.976 | 16.712 |
| 1.9 | 14.976 | 19.468 | 18.518 | 19.640 |
| 2.0 | 17.352 | 22.791 | 21.832 | 23.561 |
| 2.1 | 20.401 | 26.727 | 25.885 | 28.445 |
| 2.2 | 24.408 | 31.528 | 30.922 | 34.718 |
| 2.3 | 29.341 | 37.219 | 36.856 | 42.230 |

**Table II.** Neutron matter binding energy per neutron (in MeV) as a function of Fermi momentum, $k_F$, for the CD-Bonn potential, as obtained by BHF calculation for different single-particle potentials.

| $k_F$ (fm$^{-1}$) | E/A SPE 1 | SPE 2 | SPE 3 | SPE 4 |
|---|---|---|---|---|
| 1.0 | 5.244 | 5.722 | 5.306 | 5.307 |
| 1.1 | 6.155 | 6.649 | 6.269 | 6.276 |
| 1.2 | 7.031 | 7.594 | 7.197 | 7.209 |
| 1.3 | 7.823 | 8.553 | 8.078 | 8.101 |
| 1.36 | 8.302 | 9.166 | 8.633 | 8.665 |
| 1.4 | 8.617 | 9.589 | 9.015 | 9.058 |
| 1.5 | 9.544 | 10.778 | 10.123 | 10.194 |
| 1.6 | 10.652 | 12.152 | 11.468 | 11.582 |
| 1.7 | 11.880 | 13.774 | 13.010 | 13.190 |
| 1.8 | 13.249 | 15.615 | 14.786 | 15.068 |
| 1.9 | 14.786 | 17.816 | 16.843 | 17.272 |
| 2.0 | 16.855 | 20.580 | 19.525 | 20.189 |
| 2.1 | 19.470 | 23.854 | 22.821 | 23.803 |
| 2.2 | 22.814 | 27.814 | 26.899 | 28.342 |
| 2.3 | 26.795 | 32.463 | 31.666 | 33.690 |

From Figs. 1 and 2, one can see that the local interaction Argonne V18 is stiffer than the non-local CD-Bonn potential. It is seen that there is a weak dependence on the choice of the auxiliary single-particle potential specially at low densities. The continuous choice tends to give a slightly larger binding energies than the gap choice with a maximum deviation of about 2 MeV. The density is much less than that obtained in the case of symmetric nuclear matter (4–6 MeV).[20] This result is in agreement with that in Ref. 18. According to the results of the present work, the BHF approximation in the continuous choice can be considered a reasonable approximation up to densities that are relevant for neutron star studies. The results are also presented in Tables I and II.

The differences between the various energies are smaller in neutron matter. This is mainly due to the absence of the $^3S_1$–$^3D_1$ contribution. In pure neutron matter the strong nuclear tensor force contribution of the $T = 0$ channel is absent. The important, $^3S_1$–$^3D_1$ channel does not contribute to the energy per particle, therefore, the difference between the various potentials is expected to be small. This is very similar to the case reported by Li et al.[41] in their neutron matter calculation.

In Table III, we compare the contributions of various partial wave channels to the potential energy of PNM for different single-particle energies at a Fermi momentum $k_F = 1.8$ fm$^{-1}$. By comparing the results using the four different choices of the auxiliary single-particle potential, we note that the discrepancy between the total binding energies calculated mainly comes from the S and P channels. In the case of symmetric nuclear matter, the discrepancy between the total binding energies mainly comes from the $^3S_1$–$^3D_1$ channel.[20]

Figure 3 shows the binding energy per neutron as a function of density. The final potential appears more repulsive at a high Fermi momentum for the Argonne V18 potential than for the CD-Bonn potential. At lower densities of up to about $k_F = 1.5$ fm$^{-1}$, the two potentials produce very close binding energies per neutron, which reflects the fact that all potentials yield the same on-shell *T*-matrix. At higher densities, the quenching mechanisms due to the Pauli operator and the energy denominator [second term in Bethe–Goldstone equation (1)] account for the differences in binding exhibited by the two potentials. The potential producing the weakest tensor force (CD-Bonn) is the most attractive.

Table III. Partial-wave contributions of pure neutron matter with Argonne V18 and CD-Bonn potentials for different single-particle energies at Fermi momentum $k_F = 1.8\,\text{fm}^{-1}$. Units are given in MeV.

| Channel | BHF | | | |
|---|---|---|---|---|
| | SPE 1 | SPE 2 | SPE 3 | SPE 4 |
| Argonne V18 | | | | |
| $^1S_0$ | −19.77 | −18.19 | 18.3 | −17.74 |
| $^3P_0$ | −4.03 | −3.78 | −3.84 | −3.82 |
| $^3P_1$ | 17.5 | 18.58 | 18.16 | 18.22 |
| $^1D_2$ | −5.74 | −5.69 | −5.71 | −5.71 |
| $^3F_2$ | −1.19 | −1.18 | −1.18 | −1.18 |
| $^3P_2$ | −14.68 | −14.08 | −14.32 | −14.24 |
| $^3F_3$ | 2.82 | 2.82 | 2.82 | 2.83 |
| $^1G_4$ | −1.03 | −1.03 | −1.03 | −1.03 |
| $^3H_4$ | −0.19 | −0.19 | −0.19 | −0.19 |
| $^3F_4$ | −1.23 | −1.21 | −1.22 | −1.22 |
| E/A | 13.25 | 16.829 | 15.976 | 16.712 |
| CD-Bonn | | | | |
| $^1S_0$ | −19.65 | −19.15 | 19.21 | −19.09 |
| $^3P_0$ | −3.99 | −3.77 | −3.83 | −3.81 |
| $^3P_1$ | 17.69 | 18.65 | 18.23 | 18.26 |
| $^1D_2$ | −5.85 | −5.79 | −5.82 | −5.81 |
| $^3F_2$ | −1.18 | −1.18 | −1.18 | −1.18 |
| $^3P_2$ | −14.79 | −14.20 | −14.44 | −14.33 |
| $^3F_3$ | 2.93 | 2.93 | 2.93 | 2.93 |
| $^1G_4$ | −1.05 | −1.05 | −1.05 | −1.05 |
| $^3H_4$ | −0.26 | −0.26 | −0.26 | −0.26 |
| $^3F_4$ | −1.33 | −1.31 | −1.32 | −1.32 |
| E/A | 13.249 | 15.615 | 14.786 | 15.068 |

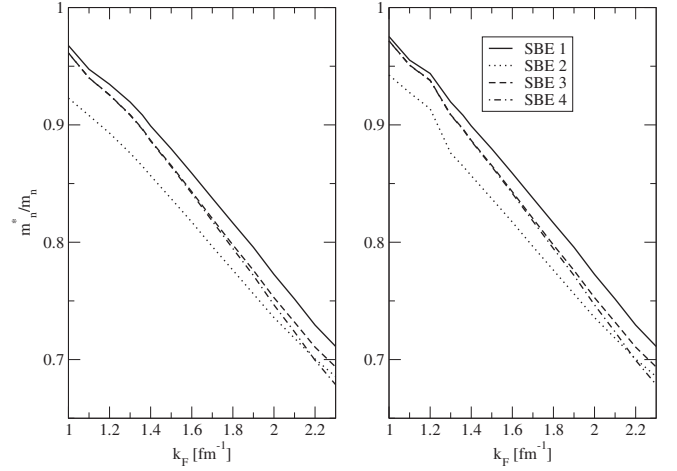

Fig. 4. Calculated effective mass for neutron matter as a function of the Fermi momentum. The left panel shows the result of the BHF calculations using Argonne potential for different single-particle energies, and the right panel is that using the CD-Bonn potential.

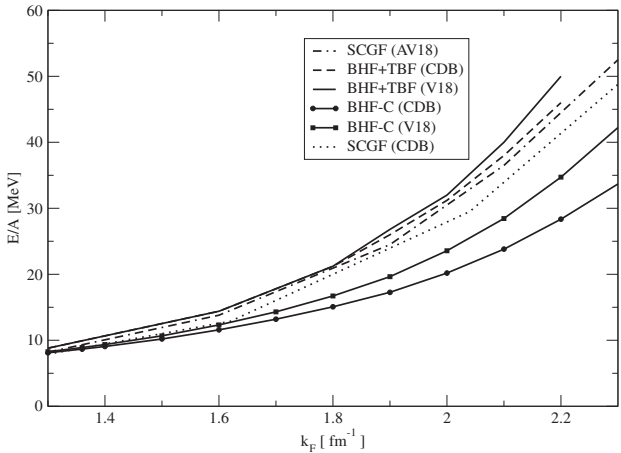

Fig. 3. Energy per particle for pure neutron matter. For comparison, nonrelativistic Brueckner calculations with TBF's are also reported.[42]

In order to establish the importance of the hole–hole term in the calculated pure neutron matter (PNM) we compared BHF calculations (which ignore the hole–hole term) with the self-consistent Green's function (SCGF), which includes the hole–hole term. In the low-density limit, BHF and SCGF coincide. As the density increases, the phase space for hole–hole propagation is no longer negligible, resulting in an enhanced repulsive effect on the total energy.

For comparison, the binding energy per neutron estimated from BHF calculation with TBF's is also reported.[42] The TBF effect is fairly small at a low Fermi momentum. As the Fermi momentum increases, the repulsive contribution of the TBF increases rapidly.

Figure 4 shows the values of the effective mass vs Fermi momentum calculated in the four cases of the single-particle energies. In the four cases, the effective masses are monotonically decreasing functions of density. The continuous choice leads to a significant reduction in the effective mass, especially at a high Fermi momentum. At a low Fermi momentum, the difference between the gap choice (SPE 2) and the other cases is shown to be large. It is well known that including second-order "rearrangement" contributions to the single-particle potential increases the theoretical $m_n^*$ values,[43] as recent perturbative calculations show.[44]

## 4. Conclusions

In this study we have established the EOS of PNM at zero temperature on the basis of the BHF and SCFG approximations. The calculations are performed for the Argonne V18 and CD-Bonn potentials up to the Fermi momentum $k_F = 2.3\,\text{fm}^{-1}$.

We presented a comparison of the energy per nucleon for neutron matter in the BHF, SCGF, and BHF+TBF approaches. We found similar binding energies at low densities, but a stiffer equation of state for BHF+TBF and SCGF calculations at higher densities. Also, it is found that the EOS is very sensitive to any change in the single-particle potential specially at higher densities.

The contribution of the hole–hole terms is repulsive, which leads to larger energies for SCGF than for BHF for all Fermi momenta in pure neutron matter. This repulsive effect is stronger in symmetric nuclear matter than in pure neutron matter. This means that the contribution of ladder diagrams is larger in the proton–neutron interaction than in the neutron–neutron interaction.

The resulting equation of state for neutron matter is in good agreement with advanced many-body calculations over a large density range. Also, according to the results of the present work, the BHF approximation in the continuous choice can be considered for a reasonable approximation up to densities relevant for neutron star studies.[45]

## Acknowledgment

One of us (Kh. Gad) would like to thank Professor Dr. H. Müther for helpful discussions.

*khalafgad@yahoo.com